\documentclass[]{aa} 
\usepackage{multirow}
\usepackage[normalem]{ulem}

\usepackage{graphicx}
\usepackage{stfloats}
\usepackage{txfonts}
\usepackage{footmisc}
\usepackage{microtype}

\usepackage{capt-of}

\usepackage[colorlinks=true,linkcolor=blue,citecolor=blue,urlcolor=blue]{hyperref}

\begin{document}

\title{Center-to-limb variations of solar active regions}
\subtitle{Observations of spots, faculae, and network in the 6173~\AA\ continuum }

 \author{A.G.M.\ Pietrow\inst{1} 
 \and S. Sumra\inst{1,2}
 \and D.J.M. Petit dit de la Roche\inst{1} 
 \and E. J. Lößnitz \inst{1,2}
 \and C. Denker \inst{1} 
 \and M. Aßmus \inst{1,3}
     }

 \institute{\inst{1}Leibniz-Institut für Astrophysik Potsdam (AIP), An der Sternwarte 16, 14482 Potsdam, Germany\\
 \inst{2}Universität Potsdam, Institut für Physik und Astronomie, Karl-Liebknecht-Straße 24/25, 14476 Potsdam, Germany\\
 \inst{3}Evangelisches Gymnasium Kleinmachnow, Schwarzer Weg 7, 14532 Kleinmachnow, Germany\\
       \email{apietrow@aip.de}\\
      }

\date{Draft: compiled on \today}

\abstract{Accurate modeling of stellar active regions (ARs) remains a major bottleneck for radial-velocity and transmission-spectroscopy studies aiming to find Earth-like planets. While much effort has been devoted to AR modeling, their center-to-limb variations (CLV) have been largely overlooked. We take a step toward remedying this by measuring the CLV of the 6173~\AA\ continuum intensity for sunspots (the whole spot,  and separate umbrae and penumbrae), faculae, network, and the quiet Sun using the Helioseismic and Magnetic Imager (HMI) onboard the Solar Dynamics Observatory (SDO). This study is based on four simple round $\alpha$-sunspots and their surroundings, as well as one strongly evolving region. After correcting for stray light, we find that relative to the quiet Sun, all components except for the umbra display reduced darkening towards the limb. Additionally, strongly evolving active regions do not appear to display significantly altered CLV profiles compared to stable active regions. Faculae and network show contrast enhancements that peak near $\mu \approx 0.3$ before declining toward the limb, reaching maxima of approximately 4\% and 2\% respectively in contrast excess relative to the quiet Sun, while the spot-to-quiet-Sun contrast rises to approximately 15\% near the limb. For both types of AR, this change in CLV behavior near the limb is likely related to the three-dimensional structure of the active regions and the rapidly changing viewing geometry. This behavior is not captured by synthetic CLVs based on PHOENIX and ATLAS model atmospheres with solar values and a different effective temperature, underscoring the need for more realistic treatments of stellar activity.}


 \maketitle
%

\section{Introduction}

The detection and atmospheric characterization of Earth-like exoplanets via radial-velocity (RV) and transmission spectroscopy methods requires an accurate understanding of the host star's surface features. Stellar active regions (AR) in particular can introduce systematic errors comparable to the planet's signal \citep[e.g.,][]{Palle2025}. These distortions depend not only on temperature and filling factor, but also on distance from the disk center. However, many studies approximate AR spectra and center-to-limb-variations (CLVs) by adjusting the temperature of their stellar models, or implicitly assume that the CLV of ARs is equivalent to that of the quiet photosphere \citep[e.g.,][]{Chakraborty2024, Adebali2026}. 

In solar physics, the CLV of ARs has been studied since the mid-20th century, though earlier qualitative mentions exist \citep[e.g.,][]{Waldmeier1939}. The first quantitative spot CLV studies were by \citet{Michard1953} and \citet{Makita1960}, who reported that in the visible the penumbral CLV closely follows that of the quiet Sun, while the umbral-to-quiet-Sun intensity ratio increases toward the limb, indicating a flatter or reduced darkening towards the limb of the umbra relative to the surrounding photosphere. This was also confirmed by \citet{Rodenberg1966}, \citet{Mattig1969}, and \citet{Fay1972}. \citet{Albregtsen1984} studied this effect over a solar cycle and found changes in umbral contrast but not in the CLV. 

Faculae, which we define as the photospheric regions under plage that appear bright near the limb \citep{Cretignier2024}, were found by \citet{Foukal1991} and \citet{Kotov1994} to have a brightening relative to the quiet Sun that is largely wavelength-independent in the visible, which is in line with a geometric hot-wall effect \citep{Spruit1976}. A similar limb-enhanced relative brightening has been observed for plage in the chromospheric \ion{Ca}{II}\,K line \citep{Cretignier2024} and even in transition region diagnostics \citep{Kayshap2024}, and has been confirmed through stereoscopic observations \citep{Albert2023, Albert2026}. The CLV of the network, which we define as a weaker and less concentrated form of faculae without overlying plage, was investigated by \citet{Auffret1991}, who found that the contrast increases from disk center toward the limb, reaching a maximum near $\mu = 0.3$, before declining again. This is in line with the results of \citet{Yeo2013} and \citet{Criscuoli2017}, who studied the CLV of the combination of faculae and network, segmenting by magnetic field strength and found a similar peak which shifted towards $\mu=0$ with stronger magnetic fields.

\begin{figure*}[!b]
    \centering
    \includegraphics[width=1\linewidth]{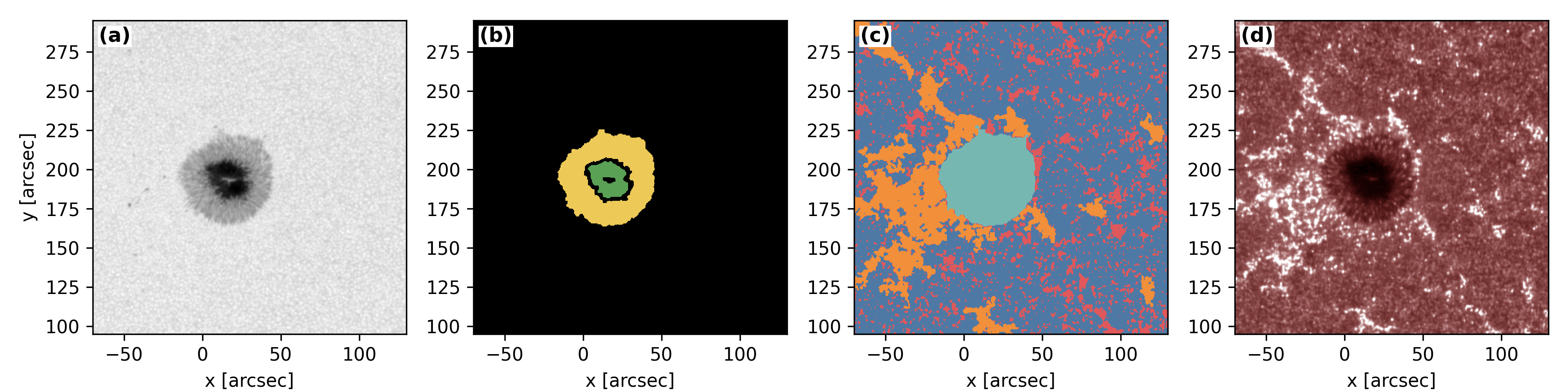}
    \caption{HMI continuum and AIA 1700~\AA\ observations of active region NOAA~12738 near disk center, showing the segmentation process. \textbf{(a)} HMI continuum intensity, \textbf{(b)} HMI segmentation mask selecting the umbra (green) and penumbra (gold), \textbf{(c)} AIA segmentation mask showing the full spot (teal), network (red), faculae (orange), and quiet Sun (blue), and \textbf{(d)} AIA 1700~\AA\ intensity. An animation of this figure over the course of the spot's disk passage is available online.}
    \label{fig:segmentation}
\end{figure*}

For the quiet Sun, several modern CLV studies now exist, most notably the spectral atlas by \citet{Ellwarth23}, which expands on the work of \citet{Neckel94}, as well as recent studies of chromospheric line CLVs \citep{Ginar2021, Pietrow22b, pietrow2026}. However, at this time, not much has been done on this topic with respect to ARs, despite the recent increase in interest in the topic from a stellar perspective. Most notable is \citet{Palumbo2024}, who demonstrated the impact of AR CLVs on RV measurements, but never investigated the intensity variations.

In this work, we present and quantify the CLV of spots (umbrae and penumbrae), faculae, and network, using space-based observations, and test whether current synthetic model atmospheres can reproduce the observed behavior.

\begin{table}
\centering
\caption{Active regions analyzed in this study. The first four are simple $\alpha$-spots; the last is a complex $\delta$-spot.}
\small
\label{tab:ar_dates}
\begin{tabular}{cccc}
\hline\hline
Active region & East limb & Meridian crossing & West limb\rule[-4pt]{0pt}{13pt} \\
\hline
\href{https://www.spaceweatherlive.com/en/solar-activity/region/12519.html}{NOAA 12519} & 2016-03-10 & 2016-03-13 22:00 & 2016-03-23 \rule{0pt}{8pt} \\
\href{https://www.spaceweatherlive.com/en/solar-activity/region/12526.html}{NOAA 12526} & 2016-03-25 & 2016-03-30 10:00 & 2016-04-07 \\
\href{https://www.spaceweatherlive.com/en/solar-activity/region/12670.html}{NOAA 12670} & 2017-08-03 & 2017-08-07 15:00 & 2017-08-15 \\
\href{https://www.spaceweatherlive.com/en/solar-activity/region/12738.html}{NOAA 12738} & 2019-04-08 & 2019-04-13 10:00 & 2019-04-21 \\
\hline
\href{https://www.spaceweatherlive.com/en/solar-activity/region/12673.html}{NOAA 12673} & 2017-08-30 & 2017-09-03 19:00 & 2017-09-10\rule{0pt}{8pt} \\
\hline
\end{tabular}\\\smallskip
\tiny{Active region details: \url{www.spaceweatherlive.com/en/solar-activity/region/[NOAA].html}}
\end{table}

\begin{figure*}
    \centering
    \includegraphics[width=1\linewidth]{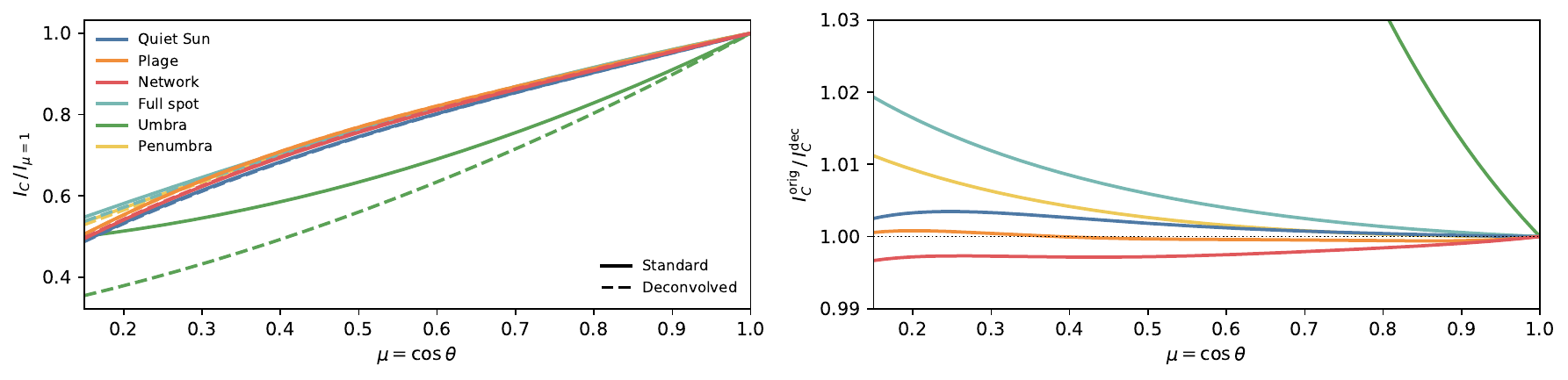}
    \caption{CLV profiles for NOAA 12738 before and after straylight correction and PSF deconvolution. Left: Normalized CLV curves for each feature type, shown before (solid) and after (dashed) deconvolution. Right: Ratio of the original to the deconvolved profiles for each feature type. }
    \label{fig:clvdecon}
\end{figure*}

\begin{figure*}
    \centering
    \includegraphics[width=1\linewidth]{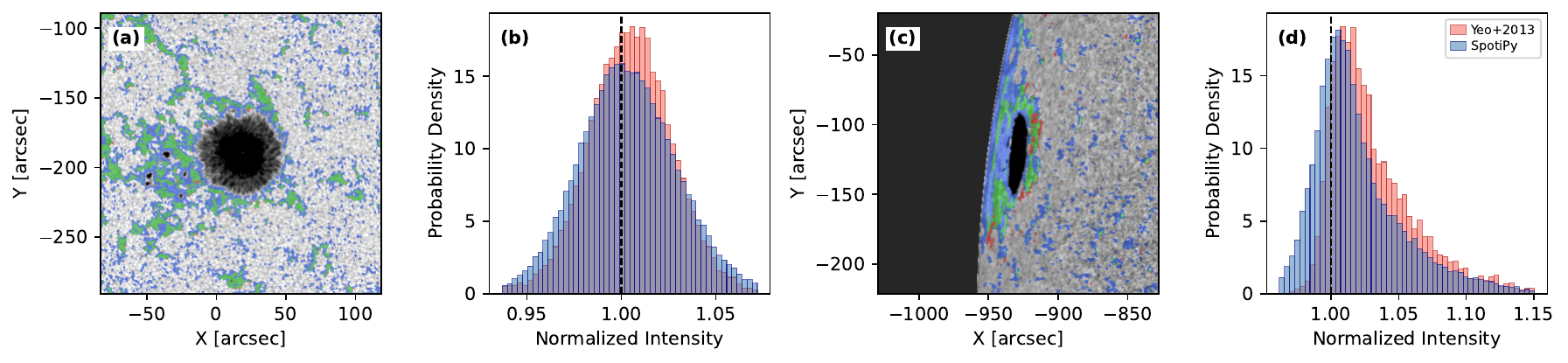}
    \caption{HMI continuum intensity maps and normalized intensity distributions for NOAA 12738 observed near disk center (a,b) and near the limb (c,d). Colored overlays in (a) and (c) indicate facular pixels identified by the \citet{Yeo2013} magnetic threshold method (red), SpotiPy (blue), and their intersection (green). Histograms in (b) and (d) show the distribution of normalized continuum intensity for pixels identified by SpotiPy (blue) and \citet{Yeo2013} (red). An animation of this figure over the course of the spot's disk passage is available online.}
    \label{fig:compareyeo}
\end{figure*}

\section{Observations and methods}

In this section, we describe the selection and tracking of AR features, as well as segmentation methods for feature identification. 

\subsection{Data selection}
To obtain the AR CLVs, these regions need to be tracked as they traverse the solar disk, populating the $\mu$-space (see video of Fig.~\ref{fig:segmentation}). This is done using the synoptic observations from the Helioseismic and Magnetic Imager \citep[HMI,][]{Scherrer2012} on board the Solar Dynamics Observatory \citep[SDO,][]{Pesnell2012}. HMI observes the Sun in six wavelength points around the \ion{Fe}{I}~6173~\AA\ line at a 45~s cadence and with a resolution of 0.5 arcseconds per pixel since its launch in 2010. One of the derived data products from these observations is the reconstructed continuum around 6173~\AA.  We focus on this continuum point, as \citet{Cohen2015} showed that the other points are not reliably sampling the line in ARs due to Zeeman broadening. 

To ensure that we primarily measure the CLV rather than evolutionary changes, we selected four isolated $\alpha$-spots \citep{Hale1919} from the catalog of \citet{Emily2025}, all classified as grade 3, meaning they remain unchanged during their disk passage and have no neighboring spots. This type of spot is also commonly simulated in both solar \citep[e.g.,][]{Rempel2009, Schmassmann2021} and stellar models \citep[e.g.,][]{Soap2012, Cauley2018, Petit2024}. 
To test the effect of evolution on such measurements, we also include a so-called $\delta$-region \citep{Kuenzel1960} that undergoes strong morphological changes and flux emergence during its disk passage \citep[see e.g.,][]{Kontogiannis2024}. 

Spots nearly always have surrounding faculae and network \citep{Solanki2003}, therefore, these can be selected in the vicinity of the tracked spots and do not have to be tracked individually. The five selected datasets and the dates of their meridian crossing are summarized in Table~\ref{tab:ar_dates}.

\subsection{Stray light correction}
When conducting CLV studies such as this one, it is important to account for stray light from the surrounding quiet Sun leaking into the sunspot umbra. This effect becomes increasingly significant toward the limb, where foreshortening reduces the projected umbral area and thus increases the relative contribution of scattered photospheric light \citep[e.g.,][]{Beck2011}. For HMI, stray light is mitigated by the pointspread function (PSF) correction algorithm of \citet{Norton2026}, following a method analogous to that of \citet{Hofmeister2025} for AIA (see Fig.~\ref{fig:clvdecon}). The effect is particularly strong for spots, where for the umbra a difference of over 20\% is reported, while for the averaged spot just over 2\%.

For each dataset, a deconvolved data product was created by the Joint Science Operations Center (JSOC) team, to which the segmentation masks were applied to extract the CLV curves.

\subsection{Segmentation}
For each timestep, a 150'' x 150''  window is drawn around the AR, which was found to be large enough to always contain all five AR types and the quiet Sun. 

Traditionally, faculae and network have been segmented based on their magnetic flux. However, the universality of this assumption has been questioned by, e.g., \citet{Criscuoli2017}, who found that flux alone is an unreliable discriminant between the two. Compounding this, magnetograms are notoriously inaccurate near the limb \citep[e.g.][]{Hoeksema2014}. For this reason, we propose an alternative form of segmentation based on the 1700~\AA\ images from the Atmospheric Imaging Assembly \citep[AIA,][]{Lemmen2012} onboard SDO, as the bright point morphology is known to trace magnetic structures everywhere on the disk \citep[see Fig.~\ref{fig:segmentation} and][]{Rutten1999, Tahtinen2022}.

This is in line with \citet{Simoes2019} who found that the bright structures in this channel form primarily in the high photosphere, but still above the formation height of the 6173~\AA continuum. This means that we expect a slightly larger footprint for faculae and network due to the expanding nature of these features \citep[e.g.][]{roberta2020}. We compare the magnetic-based segmentation method of \citet{Yeo2013} with our intensity-based method in Fig.~\ref{fig:compareyeo}. Near disk center, the two methods agree very closely with our method, sampling slightly more area, which results in a more symmetrical histogram. The histograms also largely overlap, centered around  $I_c$ as is expected for a $\mu$-angle where faculae and network are not visible. Near the limb, our threshold method seems to work similarly overall and better at the extreme limb. This is also reflected in the histogram.

After setting the thresholds, a radial average is computed for each disk and used to remove the limb darkening by dividing the disk by a 2D version of this profile, thereby constructing a limb-darkening-corrected disk, which is used to segment features. The segmentation was automated using the \texttt{Spotipy} package \citep{Emily2025}. From HMI we obtain the umbra and penumbra of sunspots, and from AIA we obtain faculae, network, and quiet Sun.  On these normalized disks, empirically optimized thresholds are applied where umbrae fall within 0.04--0.22, penumbrae 0.30--0.47, quiet Sun 0.85--1.15, and faculae/network lie above 1.20. We define bright patches of 450 pixels or more as faculae, and smaller ones as network. Similar thresholds are used in literature \citep[e.g.][]{Palumbo2024, Korda2026}. The resulting masks are then applied to the stray light corrected images to extract CLV curves.  We note that this fixed set-size difference corresponds to different projected areas for faculae across the disk. However, we found empirically that this value provides the best segmentation between the two feature types.

\begin{figure*}
    \centering
    \includegraphics[width=1\linewidth]{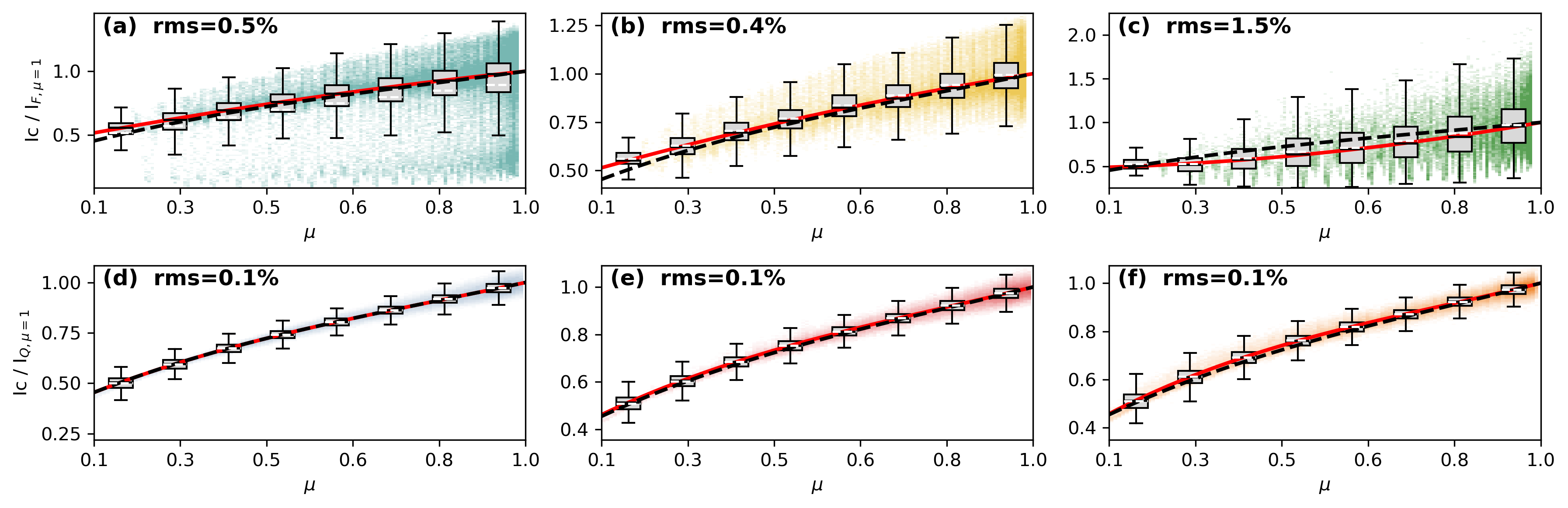}
    \caption{CLV intensity profiles for different solar surface features in the 6173~\AA\ continuum of active region NOAA~12738. Panels show the a) full sunspot, b) penumbra, c) umbra, d) quiet Sun, e) network, and f) faculae, respectively. Each panel shows the normalized intensity $I/I(\mu=1)$ as a function of $\mu$. Individual measurements are shown as a 2D histogram, while gray candle stick plots represent the full intensity distribution within bins of $\Delta\mu = 0.05$. The red curve shows a polynomial fit to the data, and the black curve indicates the corresponding quiet-Sun polynomial for reference. 
    }
    \label{fig:raw_IC}
\end{figure*}

\begin{figure*}
    \centering
    \includegraphics[width=1\linewidth]{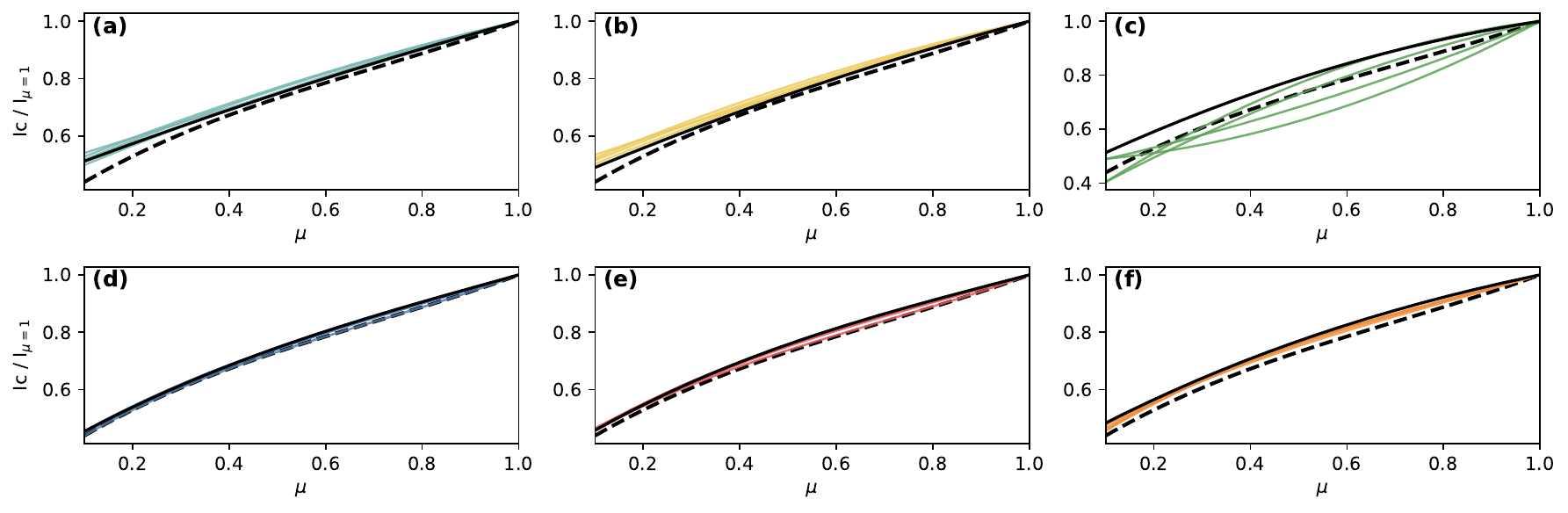}
    \caption{
    Fitted CLV intensity variations for five ARs (see Table~\ref{tab:ar_dates}) for the same surface features shown in Fig.~\ref{fig:raw_IC}. The colored curves represent regions containing stable $\alpha$-spots, while the black curve represents a region containing a $\delta$-spot that evolved strongly during its disk passage. 
    }
    \label{fig:all_IC}
\end{figure*}

\begin{table*}
\centering
\caption{Mean normalized CLV polynomial coefficients, averaged over active regions  NOAA 12526, 12738, 12519, and 12670. The normalization factor is given in the last column.}
\label{tab:clv_poly}
\begin{tabular}{l c c c c c c}
\hline\hline
Component & Order & $c_{3}$ & $c_{2}$ & $c_{1}$ & $c_{0}$ & $I_{\mu=1}/I_{\mathrm{QS},\mu=1}$ \rule[-5pt]{0pt}{15pt}\\
\hline
Quiet Sun & 3 & $+0.295 \pm 0.080$ & $-0.702 \pm 0.121$ & $+1.057 \pm 0.051$ & $+0.351 \pm 0.010$ & $1.000 \pm 0.000$ \rule{0pt}{10pt}\\
Faculae & 3 & $+0.311 \pm 0.065$ & $-0.785 \pm 0.139$ & $+1.114 \pm 0.091$ & $+0.360 \pm 0.015$ & $1.001 \pm 0.004$ \\
Network & 3 & $+0.301 \pm 0.071$ & $-0.729 \pm 0.096$ & $+1.067 \pm 0.037$ & $+0.360 \pm 0.006$ & $0.994 \pm 0.002$ \\
Full spot & 2 &   & $-0.122 \pm 0.059$ & $+0.667 \pm 0.078$ & $+0.456 \pm 0.022$ & $0.748 \pm 0.015$ \\
Umbra & 2 &   & $-0.051 \pm 0.371$ & $+0.671 \pm 0.453$ & $+0.380 \pm 0.083$ & $0.235 \pm 0.016$ \\
Penumbra & 2 &  & $-0.139 \pm 0.051$ & $+0.688 \pm 0.064$ & $+0.451 \pm 0.016$ & $0.768 \pm 0.012$ \\
\hline
\end{tabular}
\end{table*}

\begin{figure*}
    \centering
    \includegraphics[width=1\linewidth]{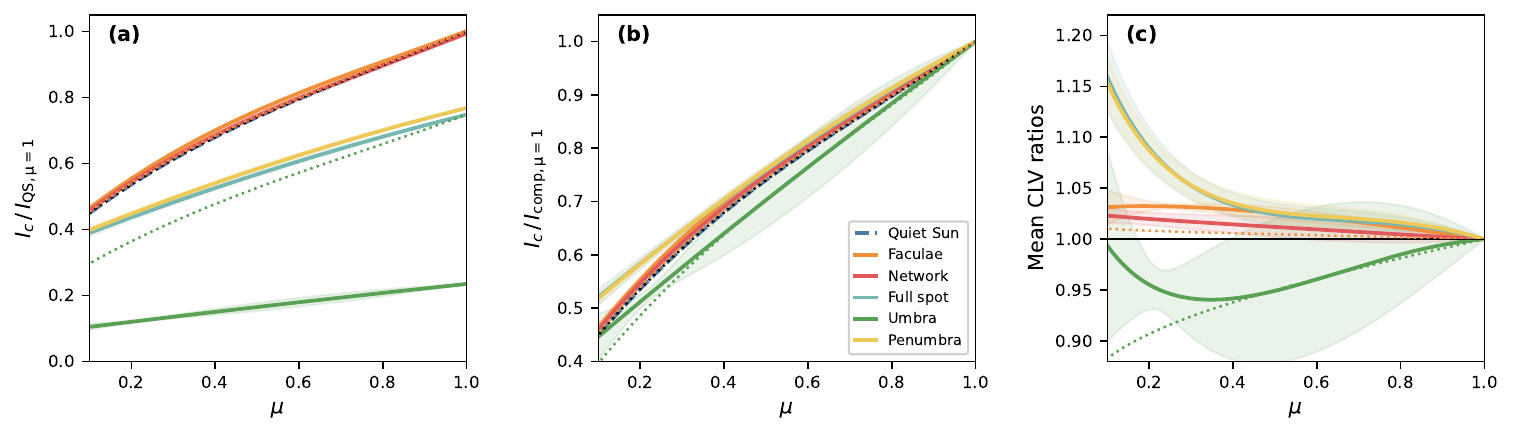}
    \caption{Mean CLV intensity variations averaged over the four $\alpha$-spots in Table~\ref{tab:ar_dates}. Uncertainties are given based on the standard deviation between each of the four averaged curves. \textbf{(a)} Mean CLV curves relative to the disk-center. \textbf{(b)} Mean CLV curves normalized to unity, quiet-Sun intensity. \textbf{(c)} Mean CLV ratios between each feature and the quiet-Sun.}
    \label{fig:mean_IC}
\end{figure*}

\section{Results and discussion}

In this section, we quantify the CLV of the 6173~\AA\ continuum intensity for umbrae, penumbrae, averaged sunspots (With an umbra penumbra ratio of around 1 to 4.), faculae, network, and quiet-Sun regions. For each of these surface features, we study the behavior of individual ARs, the averaged behavior over multiple ARs, and compare the fitted curves to the surrounding quiet Sun. The data and CLV profiles are shared in the Strasbourg astronomical Data Center (CDS) as machine-readable tables.

The intensities of the segmented data as a function of $\mu$ are given in Fig.~\ref{fig:raw_IC}. The data are then binned into eight points between $\mu$=0.95 and 0.1 with a bin size of 0.05~$\mu$. Since spots and faculae typically do not form close to the equator \citep[e.g.,][]{Solanki2003}, the value at $\mu$=1 is extrapolated. The bins are shown as candlestick plots, where the box spans the interquartile range between 25\% and 75\%, the black and white lines mark the median and mean respectively, and the whiskers extend to $2.7\sigma$.

A low-order polynomial is fitted to the data, and afterwards both the data and fit are normalized to the fitted value at $\mu$=1 to facilitate better comparisons between these regions with strongly differing intensities. The polynomial order was chosen as the lowest that reproduces the bin medians while minimizing the root mean square (RMS). A quadratic fit for the umbra and penumbra and full spot, and cubic for the rest. This process is repeated for each AR shown in Table~\ref{tab:ar_dates} and the resulting fit is shown in Fig.~\ref{fig:all_IC} in their respective colors, along with a solid black line for comparison, which represents the strongly evolving $\delta$-region. A black dashed line represents the quiet-Sun CLV. Finally, the averaged CLV of each solar surface type is shown in Fig.~\ref{fig:mean_IC}, including a shaded region which describes the standard deviation of the averaged curves. The quiet Sun CLV closely matches that of \citet{Neckel94}, with a maximum of 1.5\% mismatch near the limb.  The polynomial coefficients and their uncertainties are given in Table~\ref{tab:clv_poly}.

These empirical CLVs are then compared to synthetic curves, similarly to what was done with the quiet Sun CLV in \citet{Pietrow22b}. Synthetic CLV curves are generated from PHOENIX stellar models using the LDCU python package \citep{Husser2013, Espinoza2015}. Following \citet{Zhao2023}, a curve is generated for the quiet Sun, a spot, and a facula by assuming a $\log(g)$ of 4.4374 $\pm$ 0.0005, a solar metallicity of zero, a turbulent velocity of $1.5 \pm 0.5$~km~s$^{-1}$, and an effective temperature of $5778 \pm 50$~K, $5115 \pm 100$~K, and $5928 \pm 100$~K, respectively, for the three types of ARs.

The CLV profiles are computed over a narrow passband with a FWHM = 1~\AA\ centered at 6173\,\AA\ to match our observations. The resultant curves are sampled at 100 $\mu$-positions, with each profile being normalized to its disk-center intensity. To allow for a direct comparison, the synthetic profiles are corrected for the difference between the synthetic and observed quiet-Sun CLV by dividing by the synthetic quiet-Sun curve and multiplying by the observed one. These profiles are overplotted on the observed data in Fig.~\ref{fig:mean_IC} as dotted lines. Similar results are obtained from the ATLAS models, which were omitted for clarity. 

In all cases, except for the full spot, the whiskers encapsulate the bulk of the data points within each $\mu$-bin.  Since the penumbra dominates in both area and intensity, the full-spot CLV closely follows that of the penumbra. The full spot and penumbral CLVs are less steep than those of the quiet Sun, resulting in roughly 15\% higher intensities when normalized to their respective disk-center intensities (see Fig.~\ref{fig:mean_IC}c). This is not in line with older works, and likely the result of the stray light correction. 

The synthetic spot CLV seems to track the umbral curve until roughly $\mu=0.4$, after which they diverge, resulting in a maximum mismatch of 25\% between the empirical and model spot-to-quiet-Sun curves. This is likely caused by geometric foreshortening of the umbra near the limb as a result of the Wilson depression \citep{Wilson1774}, producing a viewing-angle-dependent effect similar to the hot-wall effect observed in faculae. Surprisingly, the $\delta$-region CLV falls within the spread of the other spots for all feature types, suggesting that spot evolution does not strongly affect the CLV when averaged over the full region. The exception is the umbra, where the region-to-region scatter is much larger, even though the reversal is present in all curves.

For faculae, we do not recover the constant increase in relative contrast toward the limb reported by earlier works. Rather, the contrast peaks near $\mu \approx 0.3$ and declines toward smaller $\mu$-values for both faculae and network. These results are in line with the behavior described by \citet{Auffret1991} for the network. The maximum contrast excess relative to the quiet Sun is approximately 4\% for faculae and 2\% for network. Our results are broadly consistent with those of \citet{Yeo2013} for combined bright regions, as they also reported a contrast peak at the same $\mu$ value position, albeit only for features with stronger magnetic field strengths and with peak contrast values approaching 10\%. The higher contrast values reported in that work are likely partially attributable to quiet Sun contamination near the limb (see Fig.~\ref{fig:compareyeo}). Finally, resolution effects likely also play a role, as prior works have reported changes in contrast with resolution \citep[e.g.,][]{Ortiz2002,Yeo2013,pietrow2020,Albert2023,Albert2026}. We therefore consider our facular and network CLV curves to represent a lower limit of the true contrast difference. Despite this, we find that the models underestimate even these contrasts and fail to reproduce the observed contrast peak, further confirming that simple temperature adjustments are insufficient to capture the CLV behavior of ARs, and that their three-dimensional structures should be accounted for.

\section{Conclusions}

Using a segmentation algorithm (see Fig.~\ref{fig:segmentation}), we measured the CLV of the 6173\,\AA\ continuum intensity for sunspots (umbrae, penumbrae, and averaged over the full spot), faculae, network, and the quiet Sun for five ARs (see Table~\ref{tab:ar_dates}), which were observed by HMI and AIA (see Fig.~\ref{fig:raw_IC}). Four of these are stable $\alpha$-type spots, while another strongly evolving $\delta$-region is included to illustrate the range of expected CLV variation within the sample. 

Average profiles for each AR type are obtained by first fitting a low-order polynomial to each individual AR dataset (see Fig.~\ref{fig:all_IC}), and then computing the mean and standard deviation of these polynomial curves across the four $\alpha$-spots (see Fig.~\ref{fig:mean_IC}).

When normalized to the quiet Sun, we find that all AR features, apart from the umbra, are flatter and thus less limb-darkened than the surrounding quiet Sun. This is consistent with earlier reports from \citet{Michard1953} and \citet{Makita1960}. The maximum contrast difference relative to the quiet Sun reaches approximately 15\% for the average spot and the penumbra and 10\% for the umbra, which is a significant discrepancy. The curves change in their behavior around $\mu$=0.4, which is likely a result of the geometrical changes of the three-dimensional structure of the spot as it approaches the limb. We note that such measurements require robust straylight corrections to truly capture the spot CLV (see Fig.~\ref{fig:clvdecon}).

The facular and network CLVs behave similarly, both showing a gradual contrast enhancement relative to the quiet Sun that peaks near $\mu \approx 0.3$ before declining again toward the limb. This is in agreement with \citet{Auffret1991}, but not with the monotonic limb-ward increase reported by \citet{Foukal1991} and \citet{Kotov1994}. The maximum contrast excess is approximately 4\% for faculae and 2\% for the network. As our segmentation method uses the AIA~1700~\AA\ channel, which samples low chromospheric structures, to generate masks for faculae and network, we expect some quiet Sun contamination at feature boundaries, and therefore consider our derived contrast values to represent a lower limit.

Model CLVs of ARs based on PHOENIX or ATLAS models created by modifying the effective temperatures fail to reproduce the observed CLV of ARs. This is likely because these models cannot reproduce the compressed stratification of sunspots \citep[e.g.,][]{Arevalo2026}, and more importantly, the three-dimensional structure of the regions. Facular approximations similarly fail to capture the aforementioned reversal caused by the hot-wall effect in faculae \citep{Spruit1976}. Surprisingly, the recovered CLVs of the regions stayed consistent even when looking at a strongly evolving and complex active region, suggesting that spot complexity does not significantly change the retrieved CLV curves. 

As the CLV is a direct observational constraint for stellar activity models used in radial-velocity and transmission spectroscopy studies, it is crucial to improve their accuracy. In future work, we aim to produce such models, as well as CLV curves at different wavelengths and in spectral lines.

\begin{acknowledgements}
AP was supported by grant PI~2102/1-1 from the Deutsche Forschungsgemeinschaft (DFG). We thank Aimee Norton, Jeneen Sommers, and Alex Koufos for making the deconvolved data available through the JSOC archive, Stefan Hofmeister for helpful discussions on the topic of deconvolution, and the anonymous referee for their comments, which greatly improved this manuscript. DeepL Write was used in copy editing of the manuscript (spelling, grammar, and readability). 
\\
\textbf{Data availability}\\
The CLV data is available at the CDS via ...
\end{acknowledgements}

\vspace{-1cm}
\bibliographystyle{aa}
\bibliography{ref}

\end{document}